\documentclass[%
 reprint,showkeys,
 amsmath,amssymb,
 aps, prl
floatfix,
]{revtex4-2}

\usepackage{graphicx}
\usepackage{dcolumn}
\usepackage{bm}
\usepackage{hyperref}
\usepackage[mathlines]{lineno}
\usepackage{color}
\usepackage[
]{geometry}

\newcommand{\mnras}{MNRAS}
\newcommand{\apjl}{ApJL}
\newcommand{\aap}{A \& A}

\newcommand{\aj}{AJ}
\newcommand{\rmxaa}{RMxAA}
\newcommand{\araa}{ARA\& A}

\newcommand{\na}{NEW ASTRON}

\newcommand{\Ms}{{\rm M}_\odot}
\newcommand{\higpus}{\texttt{HiGPUs~}}
\newcommand{\phigpu}{\texttt{PhiGPU~}}
 \newcommand{\ibh}{{\rm IMBH}}

\begin{document}

\title{Mergers of globular clusters in the Galactic disc: intermediate mass black hole coalescence and implications for gravitational waves detection
}

\author{Manuel Arca Sedda}
\email{m.arcasedda@gmail.com}
\affiliation{
Astronomisches Rechen-Institut\\
Zentrum f\"ur Astronomie der Universit\"at Heidelberg\\
69120 Heidelberg, Germany
}
\author{Alessandra Mastrobuono-Battisti}%
 \email{mastrobuono@mpia.de}
\affiliation{%
Max-Planck-Institut f\"ur Astronomie\\
K\"{o}nigsthul 17
D69117 Heidelberg, Germany\\
}%

\date{\today}

\begin{abstract}
We propose a new formation channel for intermediate mass black hole (IMBH) binaries via globular cluster collisions in the Galactic disc. Using numerical simulations, we show that the IMBHs form a tight binary that enters the gravitational waves (GWs) emission dominated regime driven by stellar interactions, and ultimately merge in $\lesssim 0.5$\,Gyr. These events are clearly audible to LISA and can be associated with electromagnetic emission during the last evolutionary stages. During their orbital evolution, the IMBHs produce runaway stars comparable with GAIA and LAMOST observations.
\end{abstract}

\maketitle

\section{Introduction}
Intermediate mass black holes (IMBHs), with masses in the range $10^2-10^5\Ms$, are thought to form via repeated star and stellar BH collisions in massive star clusters \citep{portegies00,zwart07,gaburov08,giersz15,mapelli16}. This makes globular clusters (GCs) the most promising systems to look for IMBHs \citep{silk75,maccarone08,noyola10,Lutzgendorf13,kiziltan17,mezcua17,perera2017}. The missing clear evidence of the presence of IMBHs in such systems owes mainly to the small size of their sphere of influence, namely the region of space where the IMBH dominates stellar dynamics \citep{peebles72}. The limited spatial resolution of current instruments, as well as the crowding and low luminosity of stars in GCs, makes impossible to calculate stellar radial velocities or proper motions in the IMBH immediate vicinity \cite{gebhardt05,noyola10,vandermarel10}. Moreover, several mechanisms can mimic the observational signatures typical of IMBHs, like radial anisotropies \citep{zocchi17}, observational biases like the presence of bright stars dominating the flux \citep{lanzoni13}, or the presence in the GCs centre of a population of stellar mass BHs  \citep{breen13,AS16,peuten16,abbate18,gieles18,AAG18a,AAG18b,AAG19}. The interplay between IMBHs and stars can provide a way to detect this elusive type of BHs. Close star-IMBH encounters can result in tidal disruption events \citep[TDEs][]{miller04b,shen14,Lin18}, shining-up the IMBH in multi-wavelength \citep{maccarone08,webb12,farrell14}, whereas strong scatterings can produce runaway stars \citep{pfahl05,gualandris07,fragione18b}, as the one recently observed by the LAMOST collaboration \citep{hattori19}. 

In the case of IMBH pairing, the resulting binary can appear as a low-frequency ($\sim$ mHz) gravitational waves (GWs) emitter  audible to the next generation of GW detectors \citep{eLISA13, gurkan06, amaro06,amaro09}, like LISA\footnote{\url{https://www.elisascience.org/}} \citep{eLISA13}, Taiji \citep{taiji17}, or TianQin \citep{tianqin16}. However, it is unclear what processes can favour the formation of such binaries.
In two recent papers, \cite{khopersov18} and \cite{mastrobuono19} show that massive GCs born in the MW thick disc can undergo close encounters and merge at a rate of $\sim 1.8$ per Gyr. 
In this letter we explore a new potential channel for IMBH binary formation, modelling the collision of two GCs orbiting in the Galactic disc, each hosting a central IMBH by means of direct $N$-body simulations. 
We show that the GCs merger triggers the formation of a tight binary that efficiently shrinks due to continuous interactions with cluster stars, and eventually merge releasing observable GWs. We show that IMBH mergers can be observed in future with LISA, and could produce detectable electromagnetic radiation before coalescence. We demonstrate that such mechanism can explain observations of runaway stars in the Galaxy outskirt.

\section{Method}
\label{Sec:met}

Following \cite{mastrobuono19} \citep[but see also][]{khopersov18}, we simulate the evolution of 100 GCs ($M=10^7\Ms$ each) orbiting the Galactic disc, taking into account the Galaxy as an external static potential consisting of three components: a thin and a thick disc plus a spherical dark matter halo. Each component is modelled adopting the functional forms described in \cite{allen91} assuming the parameters given in \cite{pouliasis17}, which ensure a reliable representation of the Galaxy \citep{khopersov18}.

Clusters initial conditions are taken from \cite{mastrobuono19}: 98 of them are modelled as softened Plummer spheres \citep{Plum} with scale length 20\,pc, while the remaining two clusters
are modelled with 262000 particles, corresponding to a mass resolution of $\sim 76\Ms$. 
The GCs are initialized according to two different King \citep{King} profiles, a more concentrated one with adimensional potential well $W_0 = 7$ and core radius $r_c = 1$ pc, typical for Galactic GCs, and another one with $W_0=6$ and $r_c = 4.4$ pc similar to $\omega$ Centauri, the most massive GC of the Galaxy. We assume that an IMBH sits in the centre of each cluster. The densest cluster hosts an IMBH of mass $M_\ibh = 10^4\Ms$, while for the sparser one we assume $M_\ibh = 5\times 10^3\Ms$. The resulting IMBH-to-GC mass ratio is compatible with both theoretical models \citep[]{portegies00,giersz15,AAG19} and observations \citep{Lutzgendorf13}.

We carry out the simulation up to 0.35 Gyr using two different codes: \higpus \citep{Spera,AS19}, and \phigpu \citep{berczik11,berczik13}. Both codes exploit graphic processing units parallel computing, and implement a $6^{\rm th}$-order Hermite integrator scheme with block time steps. We use \higpus to follow the first part of GCs trajectories, while \phigpu is used to follow the IMBHs binary evolution, as it implements a proper treatment for BH dynamics. We simulate part of the GCs evolution with both codes, in order to ensure their accuracy and reliability. The relative energy change at the end of the simulation is $\leq 10^{-7}$. 

\section{Results}

\subsection{Intermediate mass black hole binary formation and coalescence}
\label{Sec:IMBHB}

\begin{figure*}
    \includegraphics[width=15cm]{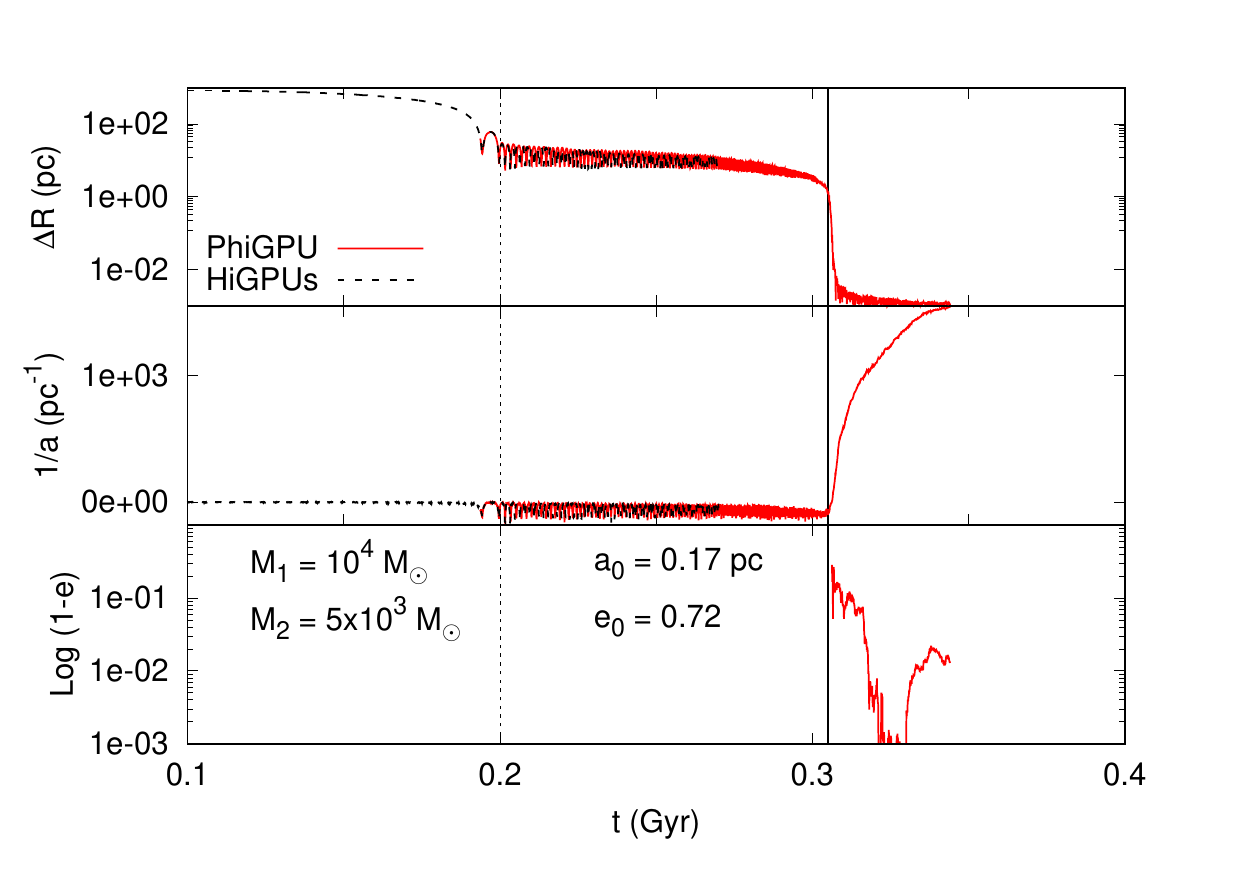}
    \caption{Time evolution of the IMBH binary orbital properties: separation (top panel), inverse semi-major axis (central panel), and eccentricity (bottom panel). Black dotted lines refer to the results obtained with the \higpus code, while red solid lines identify the \phigpu simulation. The dotted vertical line marks the moment in which the two GCs merge, while the solid
    vertical line marks the moment of the formation of the IMBH binary. Labels refers to binary initial semi-major axis ($a_0$) and eccentricity ($e_0$) values at formation.}
    \label{fig:f1}
\end{figure*}

After the GCs collision, which takes place at $\simeq 22.7$\,kpc from the Galactic centre in about 200\,Myr, the IMBHs evolution can be dissected into three distinct phases. During the first phase, which lasts $\tau_{\rm DF} \sim 100$\,Myr, the heavier IMBH settles in the centre of the cluster remnant, while the lighter one slowly inspirals to the centre due to dynamical friction. As soon as the secondary IMBH enters the primary's sphere of influence, $R_{\rm SOI} \sim 0.06$ pc, dynamical friction ceases and the two IMBHs form a binary system with initial semi-major axis $a_0 = 0.17$ pc and eccentricity $e_0 = 0.72$. The onset of binary formation is followed by a relatively short phase, $\tau_{\rm HD} \sim 30$ Myr, during which the binary hardens via stellar interactions. The hardening rate progressively slows down due to the continuous depletion of stars in the loss-cone and the binary ``stalls'', i.e. its semi-major axis varies slowly. At this stage, $t\simeq 0.3$ Gyr, the binary has $a_{\rm HD} = 6.5\times 10^{-4}$\,pc and $e_{\rm HD} = 0.98$. Figure \ref{fig:f1} shows the different phases of the IMBHs evolution. 

Despite the hardening efficiency decreases rapidly, the separation achieved at the onset of stalling is sufficiently small for gravitational waves (GWs) emission to kick in and start dominating the binary evolution, potentially leading the IMBHs to merge. Since none of the codes used to perform the simulation take into account general relativistic terms, we record $(a_{\rm HD},e_{\rm HD})$ values at the last simulated snapshot and use the analytic treatment described in \cite{peters64} to follow the binary evolution down to the merger. We find that the two IMBHs merge in $\sim 677$ Myr, thus implying a total time-span from GCs merging to the IMBHs merger $\tau_{\rm mer} \lesssim 1$ Gyr. 

Our simulation suggests that a massive ($M_{\rm bin} = 1.5\times 10^4\Ms$) IMBH binary forming after the collision between two Galactic GCs in the disc merges on a time-scale relatively short compared to typical GC ages ($\simeq 10$ Gyr). 

According to \cite{khopersov18} and \cite{mastrobuono19} estimates, we can infer for MW-like galaxy $\sim 18$ GC mergers per typical GC lifetimes. This conservative estimate assumes an initial monochromatic GC mass function, having all GCs in \cite{khopersov18} a mass of $M=10^7\Ms$. However, it must also be noted that galaxies heavier than the MW are expected to host a larger number of GCs \citep{zepf93,brodie06,hudson14}, possibly implying a larger collisions rate. 
If, as suggested in literature \citep[see for instance][]{portegies06,giersz15,AAG19}, only $10-20\%$ of GCs host an IMBH we can naively expect the formation of $\sim 2-4$ IMBH-IMBH per MW-like galaxy over a Hubble time. 

Although rare, GCs collisions in galactic discs constitute a promising way to build-up massive IMBH binaries.

\subsubsection{Gravitational waves detectability}

\begin{figure}
    \centering
    \includegraphics[width=\columnwidth]{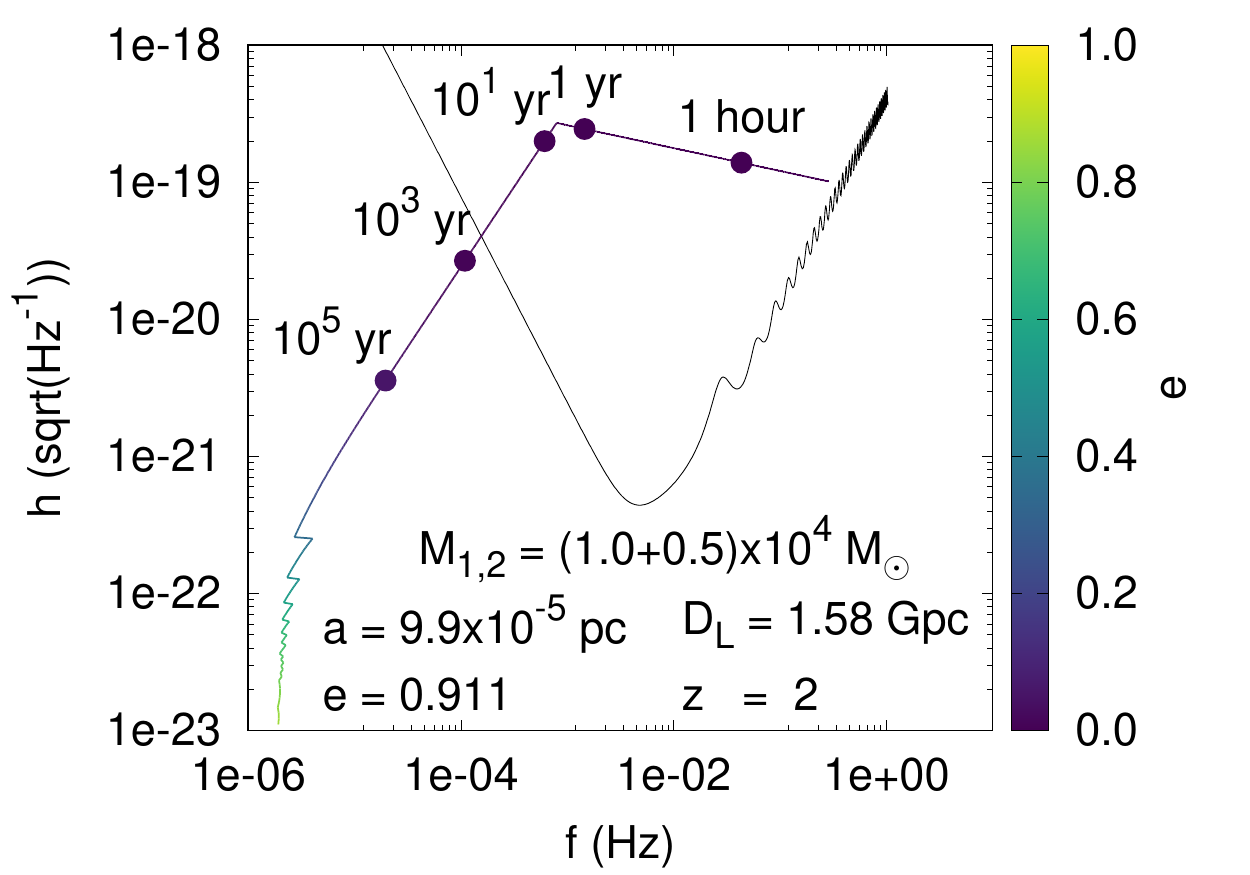}
    \caption{Amplitude of the GW signal emitted by the IMBH binary during the inspiral as a function of the leading frequency. Colour-coding identifies the binary eccentricity. Labels mark the inspiral time in different moments. The solid black line represents LISA sensitivity curve. }
    \label{fig:f2}
\end{figure}

 To check whether our IMBH binary is detectable, we compare in Figure \ref{fig:f2} the amplitude $h$ and frequency $f$ of the dominant harmonic of the binary GW signal to the LISA sensitivity curve, assuming that the source is located at redshift $z=2$, corresponding to a luminosity distance $D_L \sim 16$ Gpc. 

The source detectability depends on the LISA signal-to-noise ratio (SNR), defined as \citep{finn00,klein16,seoane17,robinson18}
\begin{equation}
\left(\frac{{\rm S}}{{\rm N}}\right)^2 = \sum_n \int \frac{h_{n}(f)}{fS_n(f)}{\rm d} \ln f,
\label{eq:SNR}
\end{equation}
where $h_{n}$ is the signal amplitude that corresponds to the $n$-th harmonic and $S_n(f)$ is the LISA averaged noise spectral density. One year prior to merger, the binary has a residual eccentricity smaller than $10^{-4}$. At these eccentricity values, the only harmonic contributing to the GW strain is the one corresponding to circular orbits, thus the sum in Equation \ref{eq:SNR} has only one term. Table \ref{tab:2} lists the SNR calculated assuming different redshifts $z$ and corresponding luminosity distances $D_L$, and further assuming that the observation begins 1\,yr prior to merger. At redshift $z<3$, the SNR will be sufficiently high to measure binary components mass with a precision of $10\%$, being ${\rm S}/{\rm R}|_{\rm threshold} = 20$ the threshold required to achieve such precision \citep{Lisa17}. 

\begin{table}
    \centering
    \caption{SNR in LISA detector at various redshifts}
    \begin{tabular}{ccc}
        \hline
        \hline
        $z$ & $D_L$ (Gpc)& SNR\\
        \hline
        $0.05$ & $0.23$ & $665$ \\
        $0.1$  & $0.48$ & $352$ \\
        $0.2$  & $1$    & $197$ \\
        $0.7$  & $4.4$  & $92$ \\
        $1.0$  & $6.8$  & $82$ \\
        $1.5$  & $11$   & $78$ \\
        $2.0$  & $16$   & $79$ \\
        $2.5$  & $21$   & $82$ \\
        $3.0$  & $26$   & $86$ \\
        \hline
    \end{tabular}
    \label{tab:2}
\end{table}

An intriguing phenomenon connected to IMBHs pairing and coalescence is the joint electromagnetic (EM) and GW emission during the binary final evolutionary stages, similarly to what expected for SMBHs \citep[see for instance][]{kocsis08}.
A condition to be fulfilled for an IMBH to sustain EM emission is to be surrounded by an accretion disc, which can trigger X-ray flares \citep{miller03,miller04,soria17,Lin18}, even jointly with radio and optical signals \citep{webb12,farrell14}, or give rise to a radio continuum source \citep{maccarone08}. 

Disrupted stars \citep{soria17,Lin18}, or leftover gas from stellar evolution \citep{abbate18} can contribute to the development of IMBH accretion discs. Furthermore, GCs moving in the Galactic disc can trap some gas from the interstellar medium \cite{li16}, which will be inevitably accreted onto the cluster heaviest object \citep{bonnell02}, most likely the IMBH. The physics of accretion discs is nowadays quite well understood for mass ranges $>10^5\Ms$, the realm of SMBHs. Hydrodynamical simulations have shown that in SMBH binaries, each component can feature a ``mini-disc'' \citep{cuadra09,roedig11,roedig14,farris14}, which sustains accretion all the way to the merging phase, making the SMBHs shine up in EM multibands throughout all the final evolutionary phases \citep{farris15,bowen18,dascoli18,zoltan17}. 

A few weeks prior to merger, the variation in the orbital evolution dominated by GW emission is expected to trigger a characteristic ``chirp''-like EM emission that can allow the identification of EM counterparts to LISA sources with masses $10^3-10^6\Ms$ at redshift $z=1-2$, and to monitor them before coalescence \citep{zoltan17}. Observing such features will be possible in the future in the optical band, through the Large Synoptic Survey Telescope (LSST\footnote{\url{www.lsst.org}}), or in X-rays through the Athena mission\footnote{\url{www.cosmos.esa.int/web/athena}}. Thus, monitoring galactic discs seeking for this kind of EM emission can allow the identification of IMBH binaries in their birth-site before the merger takes place.

\subsubsection{Properties of IMBH-IMBH binaries in galactic discs}
Binary evolution can be dissected into four stages: i) the GCs merge over a timescale $\tau_{\rm GC}$ and the heavier IMBH settles into the GC merger remnant, ii) the lighter IMBH spirals in over a dynamical friction timescale $\tau_{\rm DF}$, iii) the two IMBHs form a binary that hardens over a time $\tau_{\rm HD}$, iv) the IMBHs coalesce in a GW timescale $\tau_{\rm GW}$. Therefore, we can define a total merging time as
\begin{equation}
    \tau_{\rm mer} = \tau_{\rm GC}+\tau_{\rm DF}+\tau_{\rm HD}+\tau_{\rm GW}.
\end{equation}
The four terms in the equation above can be used to infer $\tau_{\rm mer}$ at varying IMBH masses and mass ratio. 
While the GCs collision time is independent of the IMBH mass, the dynamical friction timescale depends on the mass of the lighter IMBH $\tau_{\rm DF} \propto M_2^{-0.67}$ \citep{ASCD14a,ASCD15He}. Dynamical friction will cease when the secondary IMBH will enter the sphere of influence of the primary \citep{quinlan96}, $R_{\rm SOI} = GM_1/\sigma^2$, being $\sigma$ the velocity dispersion in the GC nucleus. The timescale over which the hardening operates scales linearly with the primary mass, being $\tau_{\rm HD} \simeq kR_{\rm SOI}$, with $k$ proportional to the binary adimensional hardening rate \citep{quinlan96}. The GW time, instead, scales with the masses of both the IMBHs as $\tau_{\rm GW} \propto \left(M_1M_2(M_{12})\right)^{-3}$ \citep{peters64}, being $M_{12}=M_1+M_2$ the binary mass. Therefore, for a generic IMBH binary with mass $M_{\rm 1n}$ and $M_{\rm 2n}$ we can write
\begin{align}
    \tau_{\rm mer,n} &= \tau_{\rm GC}+\tau_{\rm DF, n} + \tau_{\rm HD,n} + \tau_{\rm GW,n} ,\\
    \tau_{\rm DF,n} &= \tau_{\rm DF}\left(\frac{M_{\rm 2n}}{M_2}\right)^{-0.67}, \\
    \tau_{\rm HD,n} &= \tau_{\rm HD}\left(\frac{M_{\rm 1n}}{M_1}\right), \\ 
    \tau_{\rm GW,n} &= \tau_{\rm GW} \left(\frac{M_{\rm 1n}M_{\rm 2n}M_{\rm 12n}}{M_1M_2M_{12}}\right)^{-3}.
\end{align}

\begin{figure}
    \centering
    \includegraphics[width=\columnwidth]{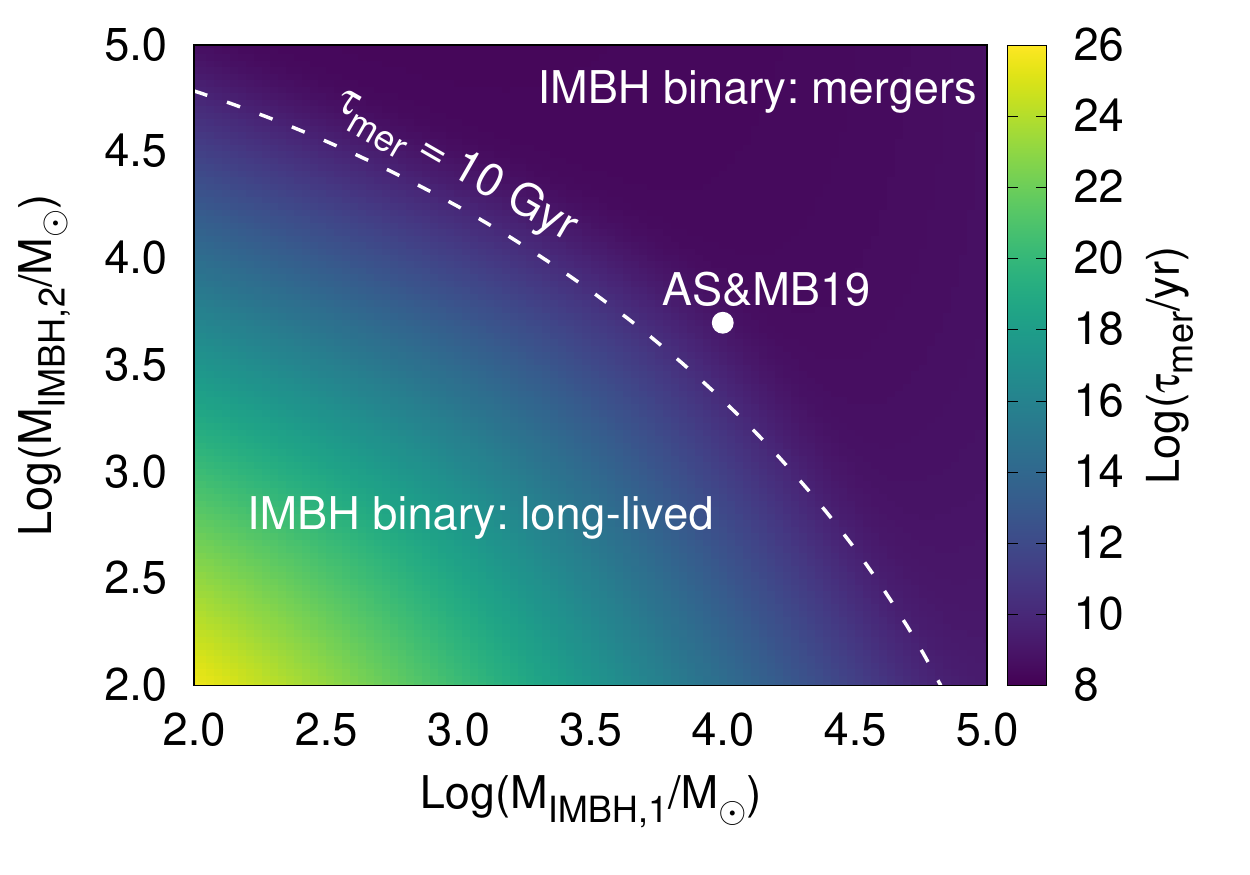}
    \caption{Total merger time at varying the primary and secondary IMBH masses. The dotted line marks the transition between binaries having merging time longer or shorter than 10 Gyr. The white dot labels our simulated model.}
    \label{fig:f3}
\end{figure}

We find a clear transition between binaries that undergo a merger within 10 Gyr, i.e. the typical age of the majority of Galactic GCs, and long-lived binaries (see Figure \ref{fig:f3}). Our predictions suggest that low-mass IMBHs, $M_{1,2}\lesssim 10^4\Ms$ are unlikely to merge within a Hubble time. 

Our simple approach suggests that a number of low-mass IMBH binaries might be lurking in the centre of GCs orbiting MW-like galactic discs. These results highlights the need for future detailed GC simulations to shed light on the probability for IMBH binaries to form following GC collisions.

\subsection{Runaway stars from star-IMBH interactions}
\label{Sec:hvs}
The IMBH binary has a crucial impact on stellar dynamics. 
For instance, close interactions with the IMBHs can trigger the ejection of stars. 
Recently, the LAMOST collaboration discovered a runaway massive star, with mass $m_* = 8.3\Ms$, traveling in the Galactic outskirts ($19$ kpc) at a velocity $\sim 568^{+19}_{-17}$ km s$^{-1}$ likely originating in a young massive cluster lurking in the Galactic Norma spiral arm \cite{hattori19}. The properties of this high-velocity star suggest that its ejection was caused by an IMBH with mass larger than $10^2\Ms$. In our simulation, we find 3542 escaper candidates after $0.4$\,Gyr, all having velocities larger than the local escape velocity, roughly $v\gtrsim 10^2$\,km s$^{-1}$. Two candidates, E25444 and E85345, are ejected from the primary IMBH after 206 and 221 Myr, with a velocity of $290$ km s$^{-1}$ and $688$ km s$^{-1}$, similar to the LAMOST star. The travel time needed to our candidates to reach such galactocentric distance is $\sim 6.9$ Myr for the fastest candidate, and $\sim 65$ Myr for the slowest one. The ejection of high-velocity stars resulting from star-IMBH interactions represents a natural outcome of the simulation, although our model was not meant to reproduce such phenomenon. Since in the simulation the ejection is triggered by the most massive IMBH, these findings might imply two possible scenarios: i) the LAMOST star ejection site still harbours the IMBH binary, ii) the ejection took place after the IMBH merger. 

\begin{table}
    \centering
    \caption{Properties of runaway stars}
    \begin{tabular}{cccc}
        \hline
        \hline
        No. & $v_{\rm ej}$ & $t_{\rm ej}$ & $t_{\rm t}$\\
         & km s $^{-1}$ & Gyr& Myr\\
        \hline
        E7110  & 2710& 0.195& 6.9\\ 
        E25444 & 290 & 0.221& 64.7\\
        E85345 & 688 & 0.206& 27.4\\
        \hline  
    \end{tabular}
    \label{tab:tab2}
\end{table}

\section{Conclusions}
\label{Sec:end}

In this letter we propose a mechanism to form IMBH binaries from the collision between two GCs in the Galactic disc. Using direct $N$-body simulations, we followed the binary formation and evolution phases, showing that the interactions with cluster stars efficiently drive the binary into the regime dominated by GWs emission, leading it to merge over a total time-span of $\sim 1$ Gyr.  
 
Our main results can be summarized as follows:
\begin{itemize}
    \item colliding GCs can trigger the formation of IMBH binaries with masses $10^2-10^5\Ms$ orbiting MW-like discs;
    \item in the case of massive IMBHs, namely $M_1+M_2\gtrsim 10^4\Ms$, the binary efficiently hardens via stellar scatterings, entering the regime of GW emission and eventually coalescing in a relatively short time-scale, $\tau_{\rm mer} \simeq 1$ Gyr,
    \item the GWs emitted during the last stages of the IMBH binary evolution are audible to LISA, being the accumulated SNR 1 yr prior to merger SNR$>80$ at redshift $z<3$,
    \item upon conservative assumptions about IMBH formation in GCs, our mechanism predicts $\sim 2-4$ IMBH mergers per MW-like galaxy per Hubble time,
    \item extrapolating our results to a wider range of IMBH masses, we find two class of IMBH binaries, either GW dominated or long-lived,
    \item the IMBHs can trigger the ejection of potential runaway stars with velocities $290-688$ km s$^{-1}$, compatible with the inferred velocity of a massive runaway star observed by the LAMOST collaboration and thus supporting the IMBH-driven ejection scenario. 
\end{itemize}

\section*{Acknowledgments}
The authors gratefully thank Peter Berczik for providing them with the \phigpu code and for his help with its compilation and execution. The authors also warmly thank Pau Amaro-Seoane for useful discussions. MAS acknowledges the Alexander von Humboldt Stiftung and the Federal Ministry for Education and Research for financial support through the research programme ``The evolution of black holes from stellar to galactic scales''. This work benefited from support from the COST ACTION CA16104 ``GW-verse'' and the International Space Science In-stitute (ISSI, Bern - Switzerland) through its International Team programme ref. no. 393 ``The Evolution of Rich Stellar Populations \& BH Binaries (2017-18)''. The authors acknowledge support by the Sonderforschungsbereich SFB 881 ``The Milky Way System'' of the German Research Foundation (DFG) and by the state of Baden-W\"urttemberg through bwHPC and the German Research Foundation (DFG), grant INST 35/1134-1 FUGG.

\bibliography{main}
\end{document}